\documentclass{article}

\usepackage{arxiv}

\usepackage[utf8]{inputenc} 
\usepackage[T1]{fontenc}    
\usepackage{hyperref}       
\usepackage{url}            
\usepackage{booktabs}       
\usepackage{amsmath,amssymb,amsfonts}
\usepackage{nicefrac}       
\usepackage{microtype}      
\usepackage{lipsum}		
\usepackage{graphicx}
\usepackage{natbib}
\usepackage{doi}

\def\diffd{\mathrm{d}}

\title{Reduction of Nonlinear Distortion in Condenser Microphones Using a Simple Post-Processing Technique}

\date{July 24, 2024}	

\author{{Petr Honzík}\thanks{This work was supported by the Grant Agency of the Czech Technical University in Prague, Grant No. SGS23/185/OHK3/3T/13 and the Institut d'Acoustique - Graduate School (IA-GS), Le Mans, France. A patent application for this invention has been filed with the French National Institute of Industrial Property (INPI) under the application number FR2404994 on May 16, 2024.} \\
	Department of Radioelectronics\\
	Faculty of Electrical Engineering\\
	Czech Technical University in Prague\\
	Technická 2, 166 27 Praha, Czech Republic \\
	\texttt{honzikp@fel.cvut.cz} \\
	\And
	{Antonin Novak}\thanks{This paper has been submitted to the IEEE Sensors Journal on July 24, 2024.} \\
	Laboratoire d'Acoustique de l'Université du Mans\\
	LAUM, UMR 6613 \\
	Institut d'Acoustique - Graduate School (IA-GS) \\
	CNRS, Le Mans Université, France \\
	\texttt{antonin.novak@univ-lemans.fr} \\
}

\renewcommand{\shorttitle}{Reduction of Nonlinear Distortion in Condenser Microphones}

\hypersetup{
pdftitle={Reduction of Nonlinear Distortion in Condenser Microphones Using a Simple Post-Processing Technique},
pdfsubject={physics.app-ph, eess.AS},
pdfauthor={A.~Novak, P.~Honzík},
pdfkeywords={Condenser microphone, Nonlinear distortion, Reduction of distortion},
}

\begin{document}
\maketitle

\begin{abstract}
					In this paper, we introduce a novel approach for effectively reducing nonlinear distortion in single back-plate condenser microphones, i.e., most MEMS microphones, studio recording condenser microphones, and laboratory measurement microphones. This simple post-processing technique can be easily integrated on an external hardware such as an analog circuit, microcontroller, audio codec, DSP unit, or within the ASIC chip in a case of MEMS microphones. It significantly reduces microphone distortion across its frequency and dynamic range. It relies on a single parameter, which can be derived from either the microphone's physical parameters or a straightforward measurement presented in this paper. An optimal estimate of this parameter achieves the best distortion reduction, whereas overestimating it never increases distortion beyond the original level. The technique was tested on a MEMS microphone. Our findings indicate that for harmonic excitation the proposed technique reduces the second harmonic by approximately 40 dB, leading to a significant reduction in the Total Harmonic Distortion (THD). The efficiency of the distortion reduction technique for more complex signals is demonstrated through two-tone and multitone experiments, where second-order intermodulation products are reduced by at least 20 dB.
\end{abstract}


\newpage

\section{Introduction}
\label{sec:introduction}

Microphones are an important part of modern technology, playing a crucial role in a variety of applications such as consumer electronics, professional audio recording and scientific measurement systems. Among the different types of microphones, condenser microphones, including MEMS (Micro-Electro-Mechanical Systems) microphones, are widely used due to their high sensitivity, wide frequency response and small size \citep{FUELDNER2020937}. However, the nonlinear distortion that can occur \citep{Yoshikawa2005} and which has an impact on audio quality and measurement accuracy is first challenging to measure \citep{Frederiksen2002,Dahlke1966} and second difficult to reduce \citep{Frederiksen1996, Fletcher2002, nikolic2023125dbspl}.

	Nonlinear distortion in microphones manifests itself as unwanted harmonic \citep{Pastille2000} and intermodulation \citep{Abuelmaatti2003} products, which affect the acquisition accuracy of the acoustic pressure signal. This distortion arises from a number of sources \citep{dessein2009modelling}. These include mechanical properties of the diaphragm \citep{Chowdhury2003}, variations in air gap damping between the membrane and backplate due to thickness changes, acoustic nonlinearities caused by high acoustic pressure in the air gap, cavity stiffness, preamplifier characteristics, and the nonlinear capacitance change of the microphone capsule \citep{Pastille2000}. Among these sources, the nonlinear capacitance change is considered the primary contributor to distortion in condenser microphones. According to Djuric \citep{djuric1976distortion}, the second harmonic, which is associated with quadratic nonlinearity of the microphone, contributes close to 90\% of the total harmonic distortion.

MEMS microphones, which have gained popularity in recent years for their integration into mobile devices and acoustic measurement systems, are inherently susceptible to nonlinear distortion due to their small size and unique construction. Previous studies have shown that the nonlinear behavior of MEMS microphones can significantly affect the accuracy of acoustic measurements, particularly at high sound pressure levels \citep{NovakAppAc2021, Printezis2024}.

Despite the advancements in microphone technology, addressing nonlinear distortion remains an important area of research. One approach to achieving a reduction of the second harmonic in single-backplate microphones is to insert a negative capacitance in the preamplifier circuit \citep{Frederiksen1996}. Fletcher \citep{Fletcher2002} suggests reducing nonlinear distortion by using a shallow parabolic backplate, although this method introduces manufacturing complications. Another technique for reducing distortion caused by the nonlinear behavior of condenser microphone capsules is the implementation of "dual backplate" technology \citep{Fuldner2015}. This method uses two backplates to reduce the nonlinear effects associated with changes in capsule capacitance, but is generally more expensive due to increased complexity and manufacturing issues. In addition, a method applicable exclusively to CMOS MEMS microphones exploits the capacitance-voltage nonlinear characteristics of a pMOS capacitor to address MEMS nonlinearity at the ASIC level \citep{nikolic2023125dbspl}.

In this context, the present study introduces a novel post-processing technique that has the potential to significantly reduce distortion levels of all single-backplate condenser type microphones\footnote{A patent application for this invention has been filed with the French National Institute of Industrial Property (INPI) under the application number FR2404994 on May 16, 2024.}. This easy-to-implement technique can be integrated into various hardware configurations, including analog circuits, microcontrollers, audio codecs, DSP units, and ASIC chips for MEMS microphones. 

The paper is structured as follows: In section \ref{sec:distortion}, we provide an overview of microphones distortion and its measurement technique used in this work. Section \ref{sec:model} elaborates on the model of the distortion in condenser microphones employed in our study. In Section \ref{sec:correction}, we present the proposed technique in detail, highlighting its simplicity and feasibility. Finally, Section \ref{sec:results} presents the measurement results, demonstrating how effectively the proposed method reduces microphone distortion.

\section{Microphone distortion measurement}	
\label{sec:distortion}

\begin{figure}[t!]
	\centering
	\includegraphics[width=7cm]{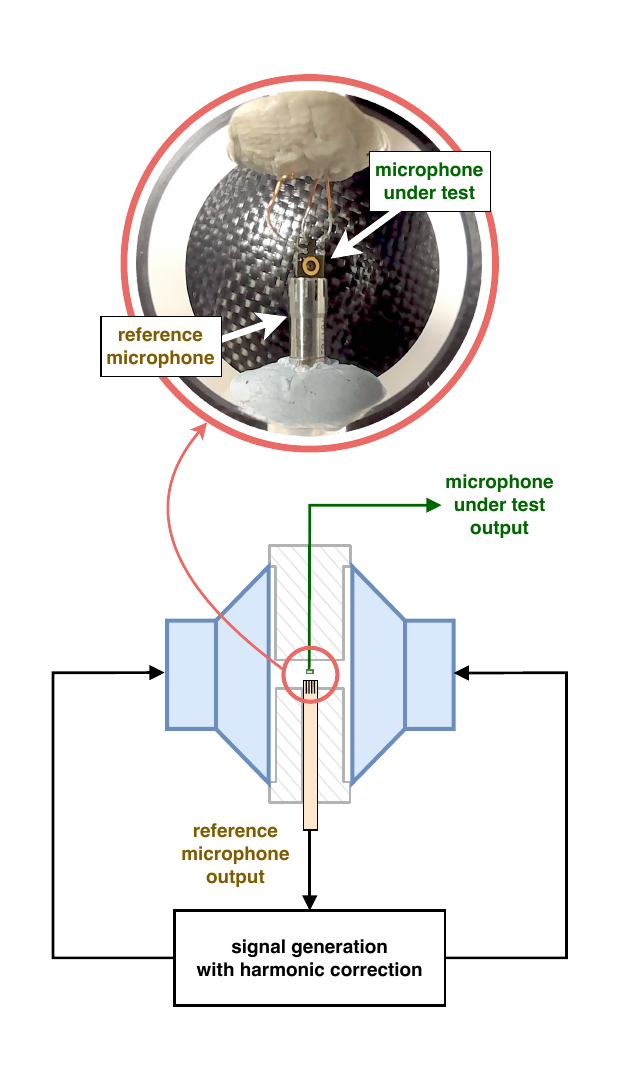}
	\caption{Schematic view of the measurement setup.}
	\label{fig:foto}
\end{figure}

\begin{figure}[t]
	\centering
	\includegraphics[width=10cm]{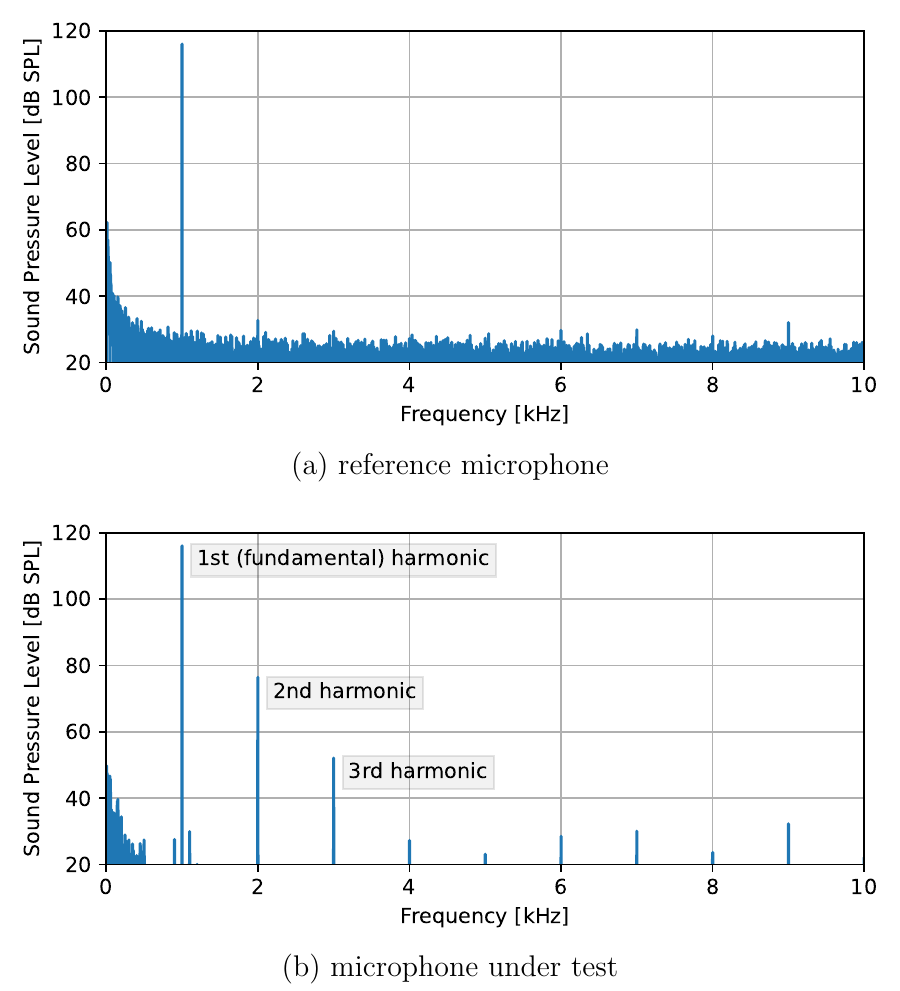}
	\caption{Example of spectra of the measured sound pressure for a 1 kHz excitation (a) with the reference microphone, (b) with the MEMS microphone under test. Note that thanks to the harmonic correction, there are no higher harmonics measured by the reference microphone.}
	\label{fig:FFT}
\end{figure}

Measuring the distortion of microphones poses significant challenges, particularly in the absence of linear sources, especially at higher sound pressure levels. For our research, we employ a recently published method enabling distortionless harmonic excitation at high-pressure levels to measure and examine the distortion of microphones \citep{Novak2018}. This method ensures a pure harmonic acoustic excitation.

The measurement setup employed in this work involves two identical 6.5" loudspeakers placed face to face, separated by a 2~cm high plastic cylindrical piece with a central aperture, with a volume of approximately $10^{-5}$ m$^3$ (see Fig.~\ref{fig:foto}). These loudspeakers are connected in parallel to create a push-push configuration, which generates pressure excitation inside the aperture with a low distortion. The MEMS microphone under test\footnote{Note, that the MEMS microphone  used in this study is an arbitrarily chosen single-backplate microphone available on the market (CUI devices, CMM-2718AB-38308-TR). Similar results and conclusions were obtained for other single-backplate MEMS microphones.} and the reference 1/8" Pressure Microphone GRAS~40DP are positioned approximately 1~mm apart within the center of the aperture.

The signal generation and acquisition are facilitated by an RME Fireface 400 sound card. To obtain a pure harmonic or pure multi-tone acoustic pressure signal inside the aperture, we use a harmonic correction technique described in \citep{Novak2018}. Consequently, the pressure signal measured with the reference microphone (distortion of which is negligible at the measured sound pressure levels) contains no unwanted higher harmonics or intermodulation products, as they are effectively suppressed to the noise level. Fig.~\ref{fig:FFT} shows an example of the spectra of the measured sound pressure for a 1~kHz excitation at a level of 116~dB SPL. The upper figure (fig.~\ref{fig:FFT}(a)) presents the efficiency of the applied harmonic correction. The higher harmonics measured by the reference microphone are suppressed to the noise level creating pure harmonic excitation inside the aperture. The lower figure (fig.~\ref{fig:FFT}(b)) shows the spectra of the sound pressure simultaneously measured by the microphone under test. This measurement is repeated for different excitation levels.

\begin{figure}[t]
	\centering
	\includegraphics[width=10cm]{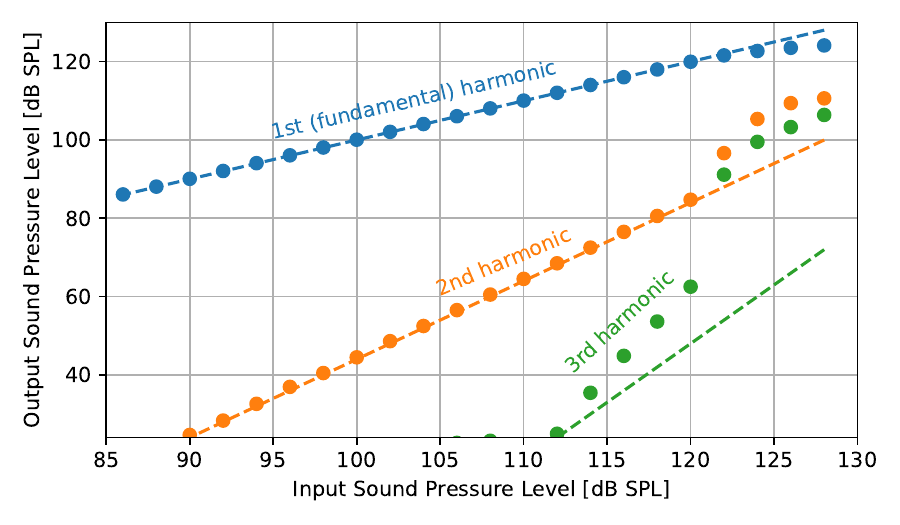}
	\caption{Sound pressure level of first three harmonics for the microphone under test as a function of reference sound pressure level, measured at 1 kHz. Measurement results are denoted by dots, model results are depicted by dashed lines.}
	\label{fig:micro_distortion}
\end{figure}

The results of such a measurement are shown in fig.~\ref{fig:micro_distortion}. The measured levels of the first harmonic (blue points), second harmonic (orange points), and third harmonic (green points) at the output of the MEMS microphone under test are shown for a 1 kHz excitation. These levels are recalculated to an equivalent input sound pressure level through the measured sensitivity of the MEMS microphone. The results are depicted as a function of sound pressure level measured by the reference microphone. The theoretical levels of the harmonic components, given by the model described below, are represented by dashed lines of corresponding colors. Up to approximately 120~dB~SPL the second harmonic fits well the predicted value, whereas the third harmonic is several dBs higher, which corresponds to previously published research \citep{NovakAppAc2021, novak2023micro}. Above 120~dB~SPL the microphone output signal presents clipping, originating probably from the electronic part of the microphone, and the measured data cannot be predicted by the model at these levels.

\section{Model of condenser microphone distortion}
\label{sec:model}


In this section, we provide a concise summary of the nonlinear model of a single-backplate condenser microphone \citep{NovakAppAc2021}. 

The relationship between the output voltage $u(t)$ and the change in capacitance $\diffd C(t)$ can be expressed as
\begin{equation}
	u(t) = - U_0 \frac{\diffd C(t)}{C},
	\label{eq:u_out}
\end{equation}
where $U_0$ represents the polarization voltage, and where the distortion is caused by nonlinear behavior of the capacitance change $\diffd C(t)$. This equation can be further expanded to the following expression:
\begin{equation}
	u(t) = U_0 \frac{C_0}{C_P + C_0} \left[\frac{\bar{\xi}(t)}{h_g} - \left(\frac{\bar{\xi}(t)}{h_g}\right)^2 \right.\\ \left.+ \left(\frac{\bar{\xi}(t)}{h_g}\right)^3 - \dots \right],
	\label{eq:u_out_nonlin}
\end{equation}
where $C_P$ represents the parasitic capacitance, $C_0$ is the static capacitance of the microphone, $h_g$ denotes the air gap thickness between the membrane and backplate, and $\bar{\xi}$ represents the mean displacement of the membrane over the backplate area. Eq.~(\ref{eq:u_out_nonlin}) can be rewritten as
\begin{equation}
	u(t) = K_0 \left[y(t) - y^2(t) + y^3(t) - \dots \right],
	\label{eq:Frederiksen}
\end{equation}
where, $K_0$ represents a constant that depends on the capacitance and polarization voltage

\begin{equation}
	K_0 = U_0 \frac{C_0}{C_P + C_0},
	\label{eq:K0_params}
\end{equation}
and $y$ represents the relative mean membrane displacement with respect to the air gap thickness. Eq.~\eqref{eq:Frederiksen} provides a simplified form to model the distortion characteristics in the electrical output of the condenser microphone.

\section{Reduction of distortion}
\label{sec:correction}
Previous research \citep{djuric1976distortion} has demonstrated that the primary distortion component in single-backplate microphones is the second harmonic. Futhermore, recent work \citep{NovakAppAc2021} has confirmed this finding for single-backplate MEMS microphones, showing that the above mentioned model accurately predicts this distortion component regardless of sound pressure level and frequency.  Consequently, the critical objective in distortion reduction is to effectively suppress mainly the quadratic component. Therefore, we can preserve the powers of $y(t)$ up to the second order in eq.~\eqref{eq:Frederiksen}, leading to
\begin{equation}
	u(t) = K_0 \left[y(t) - y^2(t)\right].
	\label{eq:u_nl_2nd}
\end{equation} 
The procedure for distortion reduction through inversion of eq.~\eqref{eq:u_nl_2nd} is described in this section.

\begin{figure}[t]
	\centering
	\includegraphics[width=10cm]{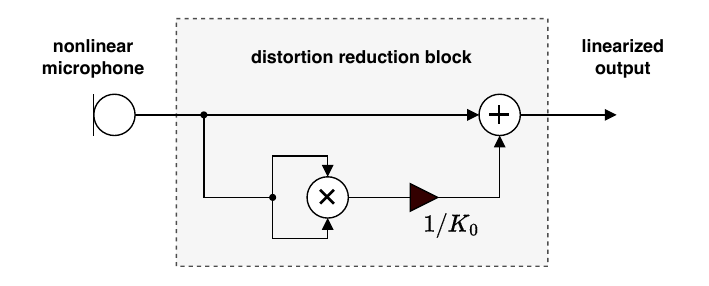}
	\caption{Block schema of the proposed distortion reduction technique.}
	\label{fig:schema_corr}
\end{figure}

First, we denote $u_{lin}(t) = K_0 y(t)$ the distortion free output signal. Then, the output voltage given by eq.~\eqref{eq:u_nl_2nd} becomes
\begin{equation}
	u(t) = u_{lin}(t) - \dfrac{1}{K_0}u_{lin}^2(t),
	\label{eq:u_nl_2nd_ulin}
\end{equation}
and one can find an inverse function to eq.~\eqref{eq:u_nl_2nd_ulin} providing an estimate of the linear output signal as
\begin{equation}
	u_{lin}(t) = \dfrac{K_0}{2}\left[1 - \sqrt{1 - \dfrac{4 u(t)}{K_0}}\right].
	\label{eq:u_lin_sqrt}
\end{equation}
Using Taylor series expansion around zero,
the estimate of the linear output signal given by eq.~\eqref{eq:u_lin_sqrt} can be further simplified to
\begin{equation}
	u_{lin}(t)  \approx  u(t) + \dfrac{1}{K_0}u^2(t).
	\label{eq:u_lin_1coef}
\end{equation}
A block schema for such a distortion reduction technique is depicted in fig.~\ref{fig:schema_corr}.

The constant $K_0$ (the only parameter necessary for the reduction of distortion using eqs.~\eqref{eq:u_lin_sqrt} or \eqref{eq:u_lin_1coef}) can be calculated from the polarization voltage $U_0$ and the static ($C_0$) and parasitic ($C_P$) capacitances of the microphone (see eq.~\eqref{eq:K0_params}). Since these parameters are not always known, a simple method for estimation of $K_0$ is proposed hereafter. 

\begin{figure}[t]
	\centering
	\includegraphics[width=10cm]{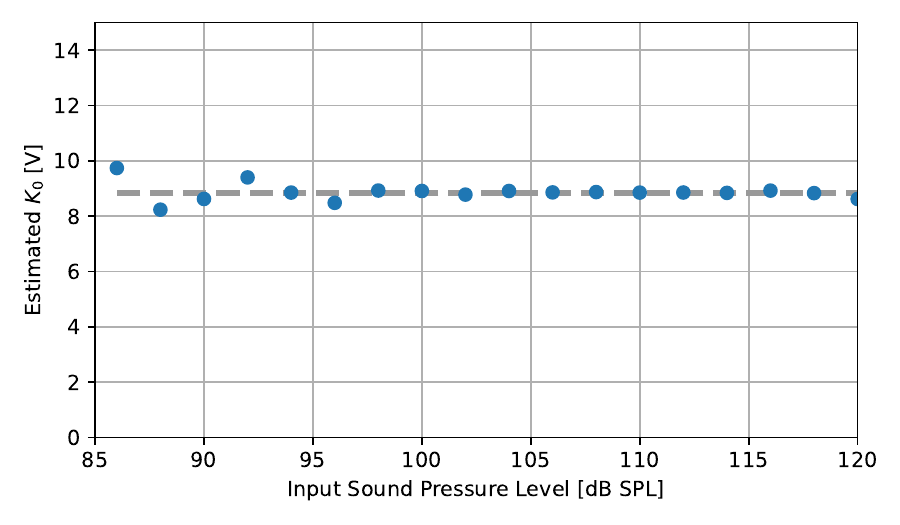}
	\caption{Estimated value of the coefficient $K_0$ as a function of reference sound pressure level. The measurement results are denoted by dots, and the selected optimal value $K_0$ is represented by the gray dashed line.}
	\label{fig:K0_estimated}
\end{figure}

Considering a pure harmonic excitation of the microphone membrane at the frequency $f_0$, the mean displacement to air gap thickness ratio can be described by $y(t)~=~y_m \sin(2 \pi f_0 t)$, where  $y_m$ denotes the amplitude. 
The absolute values of the first and second component of the Fourier series of the nonlinear output $u(t)$ \eqref{eq:u_nl_2nd} can be expressed respectively as $V_1 = K_0 y_m$, corresponding to frequency $f_0$, and $V_2 = K_0 y_m^2/2$, corresponding to frequency $2f_0$. The constant $K_0$ can be thus estimated from these measured spectral components
\begin{equation}
	K_0 = \dfrac{V_1^2}{2 V_2}.
	\label{eq:K_0estim}
\end{equation}
Such an estimate is independent of input excitation level and frequency. Fig.~\ref{fig:K0_estimated} shows the estimate of $K_0$ according to eq.~\eqref{eq:K_0estim} for different input levels at 1 kHz (blue points) using the measured data from fig.~\ref{fig:micro_distortion}, resulting in an estimated coefficient value of $K_0 = 8.85$~V (gray dashed line).

\section{Results}
\label{sec:results}

In this section the performance of the proposed distortion reduction technique is studied for different types of input signals. The impact of incorrect estimation of the parameter $K_0$ is also discussed.

\begin{figure}[t]
	\centering
	\includegraphics[width=10cm]{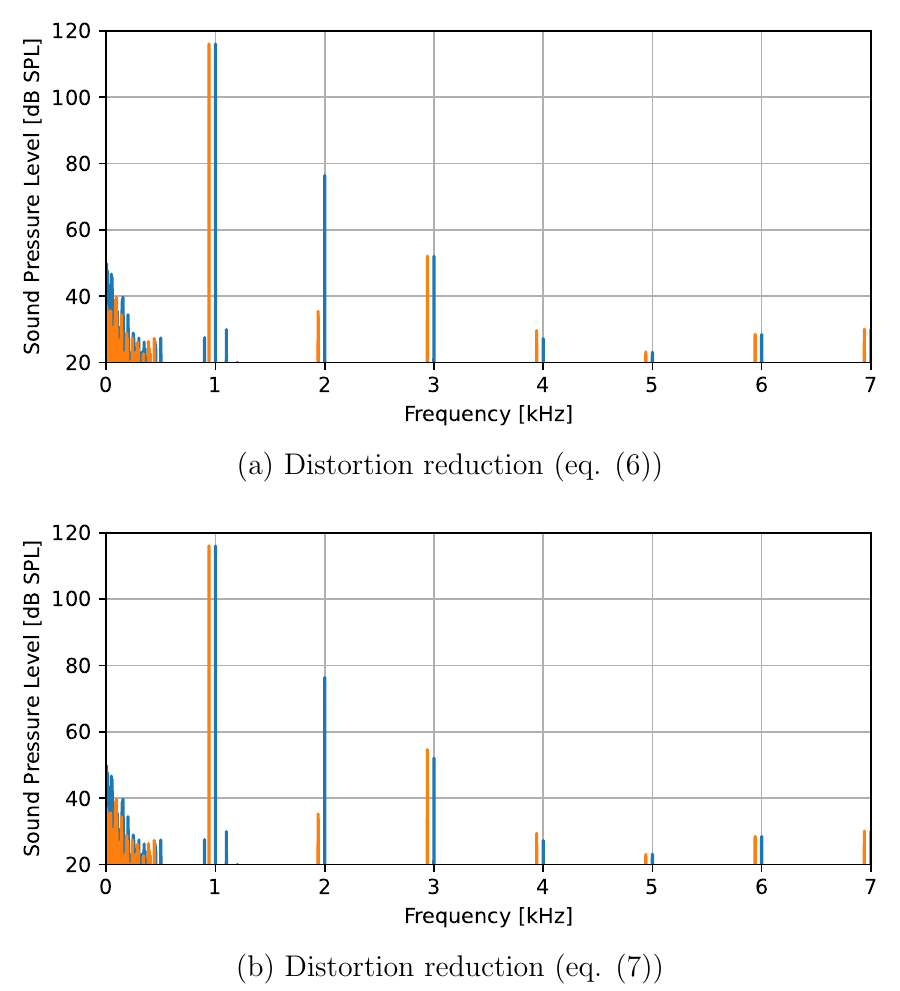}
	\caption{Spectral analysis of microphone signal for 1~kHz excitation. The blue spectrum corresponds to the original (unprocessed) signal, the orange spectrum (shifted to the left) corresponds to the signal processed by the proposed technique (Eq.~\eqref{eq:u_lin_sqrt} upper graph, Eq.~\eqref{eq:u_lin_1coef} lower graph).}
	\label{fig:correction_single}
\end{figure} 

First, the microphone was excited by a pure sine tone, which leaded to the presence of higher harmonics at its output due to the distortion, as shown in fig.~\ref{fig:correction_single} (blue line). Then, the distortion reduction technique has been applied on the microphone output signal, resulting to a significant reduction of the second harmonic component of the output signal spectrum - see orange line in fig.~\ref{fig:correction_single} (shifted to left for clarity). The upper graph of fig.~\ref{fig:correction_single} shows the result when applying the complete inverse function given by eq.~\eqref{eq:u_lin_sqrt}, the lower graph presents the result of the simplified equation \eqref{eq:u_lin_1coef}. In both cases the second harmonic component is reduced by approximately 40~dB, while the other harmonics stay almost intact (the slight increase of the third harmonic in the lower graph being negligible).  Since eq.~\eqref{eq:u_lin_1coef} provides very similar results to eq.~\eqref{eq:u_lin_sqrt} with lower computational costs, all the following results are obtained using this simple technique.


\begin{figure}[t]
	\centering
	\includegraphics[width=10cm]{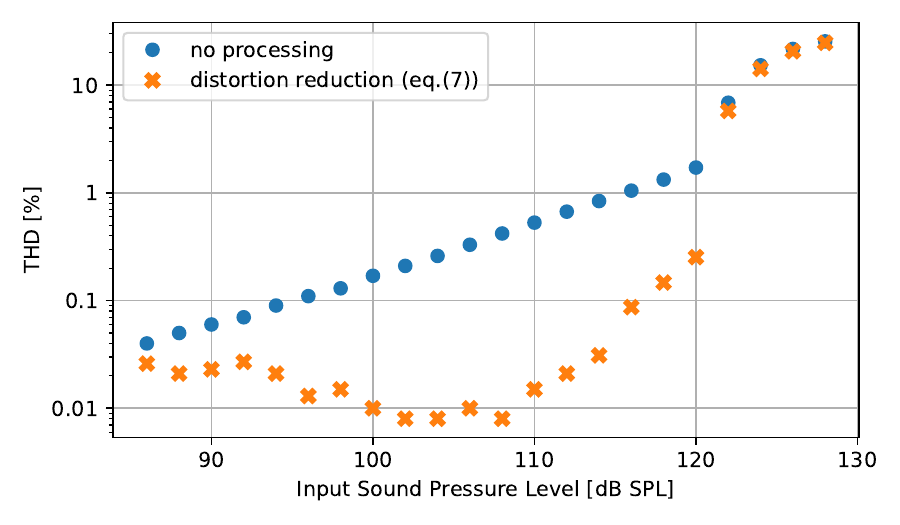}
	\caption{Total harmonic distortion as a function of reference sound pressure level. The unprocessed values are represented by blue dots, the values after applying distortion reduction are denoted by orange crosses.}
	\label{fig:correction_thd}
\end{figure}

Fig. \ref{fig:correction_thd} presents the Total Harmonic Distortion (THD) of the microphone without any postprocessing (blue points) and when applying the distortion reduction technique (orange crosses) as a function of the input sound pressure level. Except for very low levels, where the second harmonic is hidden in background noise, and very high levels, where signal clipping appears, THD is reduced 10 times near 120~dB SPL, or even 50 times near 110~dB SPL. Note that the THD obtained using the distortion reduction technique never exceeds the original THD (without any processing).

\begin{figure}[t]
	\centering
	\includegraphics[width=10cm]{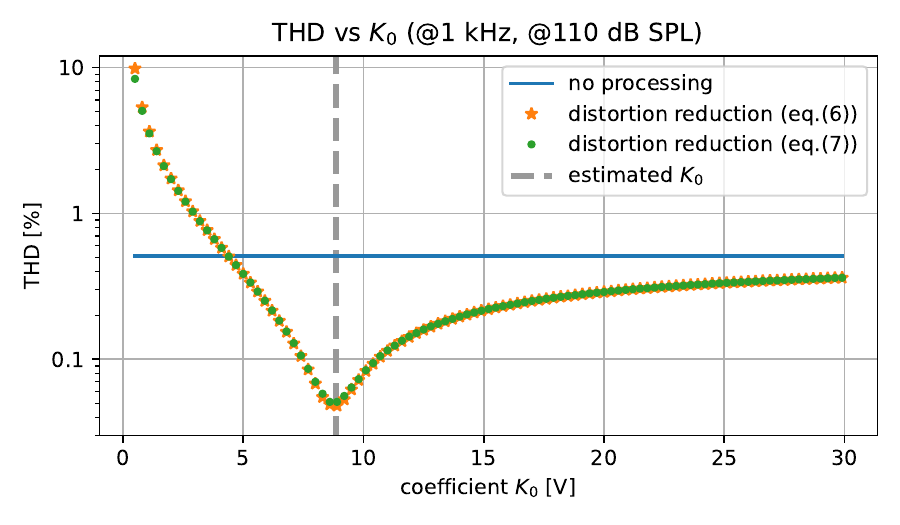}
	\caption{Total harmonic distortion of the MEMS microphone plotted as a function of the coefficient $K_0$. The unprocessed value is represented by the blue line, the technique applying the distortion reduction from Eq.~(\ref{eq:u_lin_sqrt}) by the orange stars, and the one from Eq.~(\ref{eq:u_lin_1coef}) by the green circles.}
	\label{fig:K_variation}
\end{figure}

The reduction of distortion described above was achieved using the value of $K_0$ estimated with eq.~\eqref{eq:K_0estim}. 
However, if we use eq.~\eqref{eq:K0_params}, which includes parameters $U_0$, $C_0$, and $C_P$ that may not all be known precisely, $K_0$ could be estimated inaccurately. To study the effect of uncertainty in $K_0$ on THD, we repeated the previous test for different values of $K_0$.
Fig. \ref{fig:K_variation} shows the dependence of THD on the varying value of $K_0$ using both eq.~\eqref{eq:u_lin_sqrt} (orange stars) and eq.~\eqref{eq:u_lin_1coef} (green circles) compared to THD when no distortion reduction is applied (blue line), for input sound pressure level of 110~dB~SPL. First, it is clear again that the difference between results of eq.~\eqref{eq:u_lin_sqrt} and eq.~\eqref{eq:u_lin_1coef} is negligible. Next, the estimated value of $K_0 = 8.85$~V from eq.~\eqref{eq:K_0estim} (vertical dashed gray line in  fig.~\ref{fig:K_variation}) fits well with the minimum THD value. Furthermore, fig.~\ref{fig:K_variation} shows that although the best distortion reduction is achieved with the optimal estimated value of $K_0$, THD is reduced to some extent in very large range of values of $K_0$. However, if the value of $K_0$ is largely underestimated (below 50\% of the optimal value), the THD can exceed its original value. On the other hand, overestimating the value of $K_0$ by 50\% still leads to a reduction of THD by a factor of two.  Indeed, overestimating the value of $K_0$ never causes the increase of THD above the level obtained without any distortion reduction. This is obvious from eq.~\eqref{eq:u_lin_1coef}, where the quadratic term decreases with increasing $K_0$ and the result $u_{lin}(t)$ converges to the original distorted signal $u(t)$.

\begin{figure}[t]
	\centering
	\includegraphics[width=10cm]{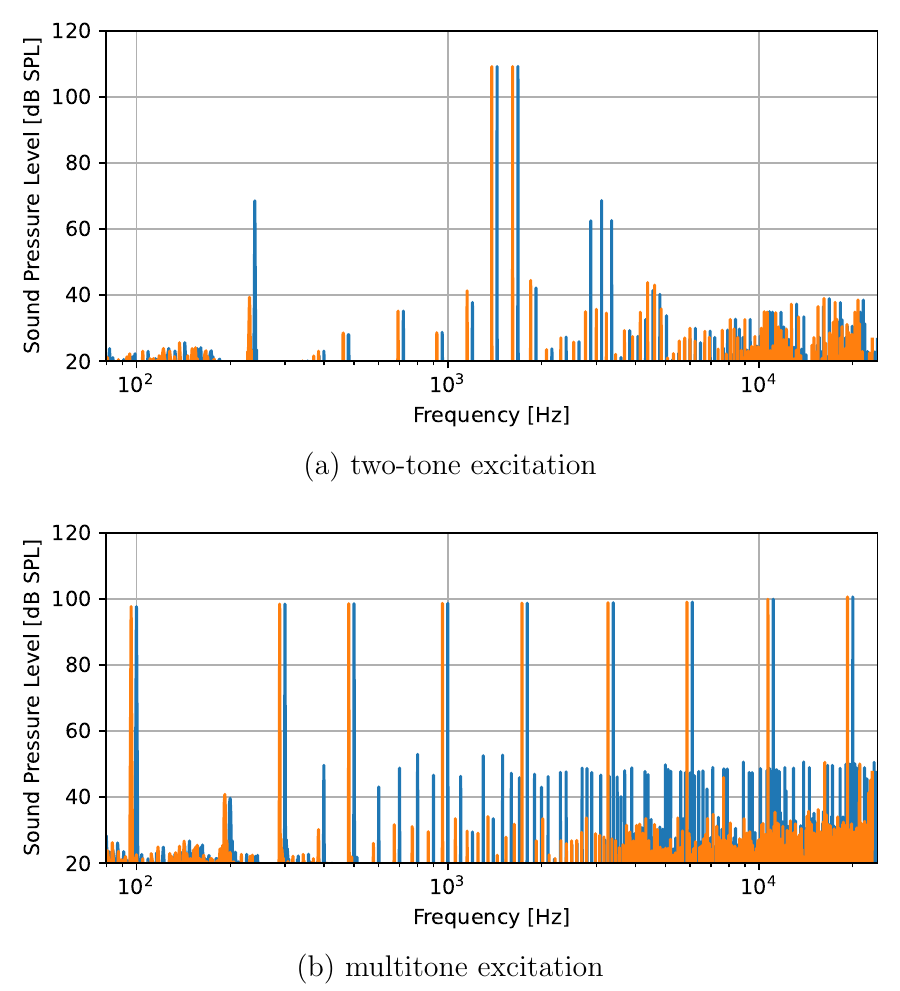}
	\caption{Spectral analysis of microphone signal for a two-tone (upper graph) and multi-tone (lower graph) excitation. The blue spectrum corresponds to the original (unprocessed) signal, the orange spectrum (shifted to the left) corresponds to the signal processed by the proposed technique.}
	\label{fig:correction_imd}
\end{figure}

In order to show the performance of the distortion reduction technique for more complex signals, a two-tone and multitone acoustic pressure signals were used to excite the microphone. Fig. \ref{fig:correction_imd} depicts the spectrum of resulting signal when applying the distortion reduction technique (orange line) and without any distortion reduction (blue line). The upper graph shows the result for the two-tone signal, where the second order intermodulation products are reduced by more than 25~dB. For the multitone signal the intermodulation products are reduced by approximately 20~dB on average, as shown in the lower graph. 

\section{Conclusion}
\label{sec:conclusion}

In conclusion, this paper presents a new technique aimed at reducing nonlinear distortion in single-backplate condenser microphones, e.g., MEMS microphones. The proposed technique offers a simple and efficient solution, requiring minimal post-processing that can be easily implemented on an ASIC chip or in external hardware (analog circuit, microcontroller, audio codec, DSP unit, etc.) and that reduces the microphone distortion within its frequency and dynamic range.

One of the key advantages of this technique is its reliance on a single parameter, $K_0$, which can be deduced either from the microphone physical parameters (polarisation voltage, static and parasitic capacitance) or from a straightforward measurement. With a correct estimation of the parameter $K_0$, the total harmonic distortion can be reduced by a factor of 10 or even 50 depending on excitation level. Remarkably, even an incorrect estimation of $K_0$ by 50\% still leads to substantial reductions in total harmonic distortion, highlighting the robustness of this approach.

The efficiency of the  distortion reduction technique of more complex signals is demonstrated through two-tone and multitone experiments. Both experiments show an important reduction of intermodulation components created by the microphone by around 20 dB. This finding further strengthens the reliability and applicability of the proposed technique.


Considering the ease of implementation, low-cost nature, and robustness of this method, it holds great promise for improving the performance of single-backplate microphones, including MEMS microphones. 



\section*{Acknowledgment}
This work was supported by the Grant Agency of the Czech Technical University in Prague, Grant No. SGS23/185/OHK3/3T/13 and the Institut d'Acoustique - Graduate School (IA-GS), Le Mans, France. A patent application for this invention has been filed with the French National Institute of Industrial Property (INPI) under the application number FR2404994 on May 16, 2024.

\bibliographystyle{unsrtnat}
\bibliography{references}  

\end{document}